\documentclass[11 pt]{article}
\usepackage[mathscr]{eucal}
\usepackage{amssymb,latexsym}
\usepackage{verbatim}
\usepackage{amsmath}
\usepackage{amsthm}
\usepackage{enumerate}
\usepackage{authblk}
\usepackage{color}
\usepackage{multicol}
\usepackage{tikz}
\usepackage{xypic}
\usepackage{caption}
\usepackage{cite}
\usepackage{hyperref}

\newtheorem{thm}{Theorem}[section]
\newtheorem{lem}[thm]{Lemma}
\newtheorem{Def}[thm]{Definition}
\newtheorem{prop}[thm]{Proposition}
\newtheorem{cor}[thm]{Corollary}

\hypersetup{
    colorlinks=true,
    linkcolor=blue,
    filecolor=magenta,      
    urlcolor=cyan,
    citecolor=black,
}

\renewcommand\({\left(}

\newcommand\wt{\widetilde}

\newcommand\e{\varepsilon}

\renewcommand\k{\kappa}

\newcommand\g{\gamma}

\renewcommand\t{\tau}

\newcommand\beq{\begin{equation}}
\newcommand\eeq{\end{equation}}
\newcommand\ben{\begin{enumerate}}
\newcommand\een{\end{enumerate}}
\newcommand\bit{\begin{itemize}}
\newcommand\eit{\end{itemize}}

\newcommand{\R}{\mathbb R}

\newcommand{\ov}{\overline}

\newcommand{\ext}{\text{{\rm ext}}}

\newcommand{\pd}{\partial}

\newcommand{\mc}{\mathcal}

\def\undertilde#1{\mathord{\vtop{\ialign{##\crcr
   $\hfil\displaystyle{#1}\hfil$\crcr\noalign{\kern1.5pt\nointerlineskip}
   $\hfil\tilde{}\hfil$\crcr\noalign{\kern1.5pt}}}}}

\newcounter{mnotecount}

\title{On the asymptotic assumptions 
for Milne-like spacetimes}

\author[1]{Eric Ling\footnote{eling@math.rutgers.edu}}
\author[2]{Annachiara Piubello \footnote{piubello@math.miami.edu}}

\affil[1]{Fields Institute, Toronto, Ontario}

\affil[2]{University of Miami, Coral Gables, Florida}

\begin{document}
\date{}
\maketitle
\vspace{.2in}

\begin{abstract}
Milne-like spacetimes are a class of hyperbolic FLRW spacetimes which admit continuous spacetime extensions through the big bang,  $\tau = 0$. The existence of the extension follows from writing the metric in conformal Minkowskian coordinates and assuming that the scale factor satisfies $a(\tau) = \tau + o(\tau^{1+\e})$ as $\tau \to 0$ for some $\e > 0$. This asymptotic assumption implies $a(\tau) = \tau + o(\tau)$. In this paper, we show that $a(\tau) = \tau + o(\tau)$ is not sufficient to achieve an extension through $\tau = 0$, but it is necessary provided its derivative converges as $\tau \to 0$. We also show that the $\e$ in $a(\tau) = \tau + o(\tau^{1+\e})$ is not necessary to achieve an extension through $\tau = 0$.
\end{abstract}


\vspace{.15in}


\section{Introduction}

Milne-like spacetimes are a class of hyperbolic  FLRW spacetimes  which admit continuous spacetime extensions through the big bang, $\tau = 0$. This extension was  observed in \cite{GalLing_con}, and further physical and mathematical properties of these spacetimes were explored in \cite{Ling_coord_sing}. (These extensions have also been noted in the physics literature, see e.g. \cite{Coleman, Nomura_Yoshida}.) The equation of state at $\tau = 0$ for a Milne-like spacetime is that of the cosmological constant, and this was generalized for nonhomogeneous versions of Milne-like spacetimes in \cite{Ling_remarks_cosmo} with applications to inflationary scenarios.

The scale factor for a Milne-like spacetime satisfies, by definition, $a(\tau) = \tau + o(\tau^{1+\e})$ as $\tau \to 0$  for some $\e > 0$. The role of the $\e$ is to achieve a continuous extension through the big bang \cite[Thm. 3.4]{Ling_coord_sing}. This extension is found by introducing coordinates $(t,x,y,z)$ where the metric in these coordinates is conformal to the Minkowski metric. In these conformal Minkowskian coordinates, a Milne-like spacetime lies within the interior of the lightcone at the origin, and the lightcone itself acts as the past boundary for the Milne-like spacetime. The asymptotic assumption $a(\tau) = \tau + o(\tau^{1+\e})$ guarantees that the metric extends continuously through the lightcone.

The asymptotic assumption $a(\tau) = \tau + o(\tau^{1+\e})$ implies $a(\tau) = \tau + o(\tau)$. In this paper, we investigate whether $a(\tau) = \tau + o(\tau)$ is necessary or sufficient to achieve an extension through the lightcone within the conformal Minkowski coordinates. We also ask whether the $\e$ in $a(\tau) = \tau + o(\tau^{1+\e})$ is necessary for an extension. We find that $a(\tau) = \tau + o(\tau)$ is not sufficient for an extension but it is necessary provided $\lim_{\tau \to 0}a'(\tau)$ exists\footnote{In this paper, $\tau \to 0$ will always mean $\tau \to 0^+$. Also, if we say a limit exists, then we always mean within the extended real number system, i.e. $\pm \infty$ are included.}. We also find that the $\e$ is not necessary. Specifically, our main theorem is the following. (We make the statements precise in the coming sections.) 

\medskip
\medskip

\begin{thm}\label{thm-main}
Let $(M,g)$ be a hyperbolic FLRW spacetime satisfying $a(\tau) \to 0$ as $\tau \to 0$ and assume that the $\tau = \text{constant}$ hyperboloids foliate $M$ all the way down to the lightcone, $\tau = 0$, within the conformal Minkowskian coordinates $(t,x,y,z)$.
\begin{itemize}
\item[\emph{(1)}] If the metric extends continuously to the lightcone, then $a(\tau) = \tau + o(\tau)$ provided $\lim_{\tau \to 0}a'(\tau)$ exists.

\item[\emph{(2)}] There are examples of scale factors satisfying $a(\tau) = \tau + o(\tau)$ such that the metric does not extend continuously to the lightcone. 

\item[\emph{(3)}] There are examples of scale factors satisfying $a(\tau) = \tau + o(\tau)$ but $a(\tau) \neq \tau + o(\tau^{1+\e})$ for any $\e > 0$ such that the metric does extend continuously to the lightcone.

\end{itemize}
\end{thm}

\medskip
\medskip

This paper is organized as follows. In section \ref{sec-review} we review preliminary material for spacetime extensions and show how Milne-like spacetimes admit such extensions through the lightcone in the conformal Minkowskian coordinates. In section  \ref{sec-necessary} we investigate the necessity of $a(\tau) = \tau + o(\tau)$ for an extension and prove (1) in Theorem \ref{thm-main} via Proposition \ref{main prop}. We also explore sufficient conditions on the scale factor that ensure $\lim_{\tau \to 0}a'(\tau)$ exists. We do this by formulating a boundary-value problem satisfied by the scale factor, and we study existence and uniqueness of solutions to this boundary value problem. At the end of section \ref{sec-necessary}, we remark what our theorems imply if one restores the speed of light $c$ in the metric and allows for arbitrary negative sectional curvature $-\k^2$ in the spatial slices; in this case, we find that $a(\tau) = c\k\tau + o(\tau)$ is necessary. In section \ref{sec-not suff} we show that $a(\tau) = \tau + o(\tau)$ is not sufficient to obtain an extension although it is sufficient to show that the $\tau =$ constant hyperboloids foliate $M$ all the way down to the light cone. We prove (2) and (3) in Theorem \ref{thm-main} via Propositions \ref{not suff prop} and \ref{eps not nec prop}, respectively.

Milne-like spacetimes were found by investigating low regularity aspects of Lorentzian geometry. This is a growing field with many tantalizing problems to solve. For low regularity causal theory, generalizations, and various results, see \cite{ChrusGrant, Leonardo, Leonardo_Soultanis, Ling_causal_theory, Minguzzi_cone, future_not_open, Clemens_GH, Lesourd_Minguzzi, Greg_Graf_Ling_AdSxS2}. For low regularity spacetime inextendibility results, see \cite{SbierskiSchwarz1, SbierskiSchwarz2, SbierskiHol, GalLing_con, GLS, GrafLing, ChrusKlinger}. For the singularity theorems in low regularity, see \cite{Hawking_Penrose_C11, Hawking_sing_low_reg, Penrose_sing_low_reg, Graf_sing_thm, Schin_Stein}. For results on geodesics and maximizing causal curves in low regularity, see \cite{Clemens_Steinbauer, Lorentz_meets_Lipschitz, Schin_Stein}. For results on Lorentzian length spaces, see \cite{Lorentzian_length_spaces, cones_as_length_spaces, length_spaces_causal_hierarchy, time_fun_on_length_spaces, Lorentzian_analogue}. Lastly, for results related to the null distance function and other notions of distance defined on a spacetime, see \cite{Null_distance, Spacetime_distances_exploration, prop_null_dist, null_distance_lorentzian_length_spaces, AnnaChristina}.

\section{Preliminaries and review of Milne-like spacetimes}\label{sec-review}

Our conventions will follow \cite[sec. 2]{Ling_coord_sing} which we briefly review. 

Let $k \geq 0$ be an integer or $\infty$. A $C^k$ \emph{spacetime} $(M,g)$ is a $C^{k +1}$ four-dimensional manifold $M$ (connected, Hausdorff, and second countable) equipped with a $C^k$ Lorentzian metric $g$ (i.e. its components $g_{\mu\nu}$ are $C^k$ functions in any coordinate system) and a time orientation induced by some $C^0$ timelike vector field. A \emph{future directed timelike curve} $\g \colon [a,b] \to M$ is a piecewise $C^1$ curve such that $\g'(t)$ is future directed timelike for all $t \in [a,b]$, including its break points and endpoints (understood as one-sided derivatives). \emph{Past directed timelike curves} are defined time-dually. 

Suppose $(M,g)$ is a $C^k$ spacetime, $(M_\ext, g_\ext)$ is a $C^0$ spacetime, and $\phi \colon M \to M_\ext$ is an isometric embedding preserving time orientations. We say $(M_\ext, g_\ext)$ is a \emph{continuous extension of $(M,g)$ with respect to $\phi$} if $\phi(M) \subset M_\ext$ is a proper subset. Henceforth, when convenient, we identify $M$ with $\phi(M)$.

 

Let $(M_\ext, g_\ext)$ be a continuous extension of a $C^k$ spacetime $(M,g)$ with respect to $\phi\colon M \to M_\ext$. The topological boundary of $M$ within $M_\ext$ is denoted by $\pd M$. A future directed timelike curve $\g \colon [a,b] \to M_\ext$ is called a \emph{future terminating timelike curve} for a point $p \in \pd M$ provided $\g(b) = p$ and $\g\big([a,b)\big) \subset M$. \emph{Past terminating timelike curves} are defined time-dually. The \emph{future} and \emph{past boundaries} of $M$ within $M_\ext$ are 
\begin{align*}
\pd^+M \,&=\, \{p \in \pd M \mid \text{there is a future terminating timelike curve for $p$}\}\\
 \pd^-M \,&=\, \{p \in \pd M \mid \text{there is a past terminating timelike curve for $p$}\}.
\end{align*}

\medskip

\noindent\emph{Example.} Let $(M,g)$ be the $C^\infty$ spacetime $M = (0,\infty) \times \R^3$ with $g = -f(t)dt^2 + dx^2 +dy^2 + dz^2$ where $f(t) = 1 + \sqrt{t}$. (The time orientation is induced by $\pd_t$.) A continuous extension, $(M_\ext, g_\ext)$, of $(M,g)$ with respect to the inclusion map  is given by $M_\ext = \R \times \R^3$ with metric $g_\ext = g$ for points in $M$ and $g_\ext = -dt^2 + dx^2 +dy^2 + dz^2$ for points in $(-\infty, 0] \times \R^3$. In this case, the past boundary $M$ within $M_\ext$ coincides with the hypersurface $t = 0$. 

\medskip

\begin{Def}\label{main def} \emph{Let $(M,g)$ be a $C^k$ spacetime and $\phi \colon M \to \phi(M) \subset \R^{4}$ be a global coordinate system (i.e. a $C^k$ diffeomorphism onto an open subset of $\R^4)$. We say that $\phi$ \emph{admits a past boundary} for $(M,g)$ if there is an open set $M_\ext \subset \R^{4}$ and a $C^0$ Lorentzian metric $g_\ext$ on $M_\ext$ such that $(M_\ext, g_\ext)$ is a continuous extension of $(M, g)$ with respect to $\phi \colon M \to M_\ext$ and $\pd^-M \neq \emptyset$. (Note we enlarged the codomain of $\phi$ from $\phi(M)$ to $M_\ext$.) }

\end{Def}

\medskip

Next we review Milne-like spacetimes and show that a certain global coordinate system for them admits a past boundary in the sense of Definition \ref{main def}. Milne-like spacetimes are a class of hyperbolic FLRW spacetimes. Recall that FLRW spacetimes model the large scale spatial isotropy of our universe  \cite{HE, ON, Wald}.

Consider the class of $C^k$ spacetimes
\[
M \,=\, (0, \tau_{\rm max}) \times \R^3 \:\:\:\: \text{ and } \:\:\:\: g \,=\, -d\tau^2 + a^2(\tau)h
\]
where $a \colon (0,\tau_{\rm max}) \to (0,\infty)$ is a $C^k$ function (called the \emph{scale factor}), $\tau_{\rm max} \in (0, \infty]$, and $(\R^3, h)$ is the simply connected hyperbolic space with constant sectional curvature $-1$. The time orientation on $(M,g)$ is induced by $\pd_\tau$. This class of spacetimes will be referred to as \emph{$C^k$ hyperbolic FLRW spacetimes}.  We are interested in those hyperbolic FLRW spacetimes which satisfy $a(\tau) \to 0$ as $\tau \to 0$. In this case, $\tau = 0$ is often referred to as the \emph{big bang}. If $a(\tau) = \tau$, then $(M,g)$ is called the \emph{Milne universe} \cite{Milne}. If $a(\tau) = \tau + o(\tau^{1+\e})$, then $(M,g)$ is called a \emph{$C^k$ Milne-like spacetime}. 
Evidently, the Milne universe is just one example of a $C^\infty$ Milne-like spacetime. 

Let $(M,g)$ be a $C^k$ hyperbolic FLRW spacetime. Hyperbolic space, $(\R^3, h)$, admits global coordinates $(y^1, y^2, y^3)$ where each $y^i$ takes values in $\R$. Let $\psi \colon M \to \psi(M) \subset \R^4$ denote the global coordinate system $\psi = (\tau, y^1, y^2, y^3)$. If $a(\tau) \to 0$ as $\tau \to 0$, then it's evident that $\psi$ does not admit a past boundary for $(M,g)$. The question we ask is:
\[
\emph{\text{Is there another global coordinate system, }} \phi, \emph{\text{ that does admit a past boundary?}}
\]

We review how Milne-like spacetimes answer this question affirmatively. Fix a $C^k$ hyperbolic FLRW spacetime $(M,g)$. The metric in comoving coordinates $(\tau, R, \theta, \varphi)$ is

\begin{equation}
g\,=\, -d\tau^2 + a^2(\tau)\big[dR^2 + \sinh^2(R)(d\theta^2 + \sin^2\theta d\varphi^2) \big].
\end{equation}
We introduce new coordinates $(t,r,\theta, \varphi)$ via 
\begin{equation}\label{t and r def}
t \,=\, b(\tau)\cosh(R) \quad \text{ and } \quad r\,=\, b(\tau)\sinh(R),
\end{equation}
 where $b\colon (0, \tau_{\rm max}) \to (0, \infty)$ is the $C^{k+1}$ increasing function given by 
\begin{equation}\label{b def}
b(\tau) = \exp\(\int_{\tau_0}^\tau \frac{1}{a(s)}ds\right)
\end{equation}
for some chosen $\tau_0 \in (0, \tau_{\rm max})$. (Note that for the Milne universe, $a(\tau) = \tau$, we obtain $b(\tau) = \tau$ when $\tau_0 = 1$.) Hence $b$ satisfies $b' = b/a$. Putting $\Omega = 1/b' = a/b$, the metric  is
\begin{align}\label{conformal metric intro eq}
g \,&=\, \Omega^2(\tau)\big[-dt^2 + dr^2 + r^2(d\theta^2 + \sin^2\theta d\varphi^2) \big] \nonumber
\\
&=\, \Omega^2(\tau)[-dt^2 + dx^2 + dy^2 + dz^2].
\end{align} 

The standard coordinates $(x,y,z)$ for $\R^3$ are related to the spherical coordinates $(r,\theta,\varphi)$ in the usual way. Thus hyperbolic FLRW spacetimes are conformal to (a subset of) Minkowski space. In eq. (\ref{conformal metric intro eq}), $\tau$ is implicitly a function of $t$ and $r$ given by
\begin{equation}\label{tau t r eq}
b^2(\tau) \,=\, t^2 - r^2. 
\end{equation}
Therefore the spacetime manifold $M$ lies within the set of points $t >r$.

We call $(t,x,y,z)$ the \emph{conformal Minkowskian coordinates}. Let $\phi \colon M \to \phi(M) \subset \R^4$ denote the global coordinate system
\begin{equation}\label{global coord}
\phi \,=\, (t,x,y,z).
\end{equation}
From eq. (\ref{tau t r eq}) it follows that $M$ coincides with the set in $\R^4$ given by
\begin{equation}\label{coincides eq}
\phi(M) \,=\, \big\{(t,x,y,z) \mid t > r = \sqrt{x^2 + y^2 + z^2} \: \text{ and } \: t^2 - r^2 < \lim_{\tau \to \tau_{\rm max}} b^2(\tau) \big\}
\end{equation}
if and only if $b(\tau) \to 0$ as $\tau \to 0$.

\begin{figure}[h]
\begin{center}
\begin{tikzpicture}[scale = .7]

\shadedraw [white](-4,2) -- (0,-2) -- (4,2);
\shadedraw [dashed, thick, blue](0,-2) -- (4,2);
\shadedraw [dashed, thick, blue](0,-2) -- (-4,2);

\draw [<->,thick] (0,-3.5) -- (0,2.25);
\draw [<->,thick] (-4.5,-2) -- (4.5,-2);

\draw (-.35,2.5) node [scale = .85] {$t$};
\draw (4.75, -2.25) node [scale = .85] {$x^i$};
\draw (-.25,-2.25) node [scale = .85] {$\mc{O}$};

\draw [->] [thick] (1.5,2.8) arc [start angle=140, end angle=180, radius=60pt];
\draw (2.0,3.25) node [scale = .85]{\small{The Milne universe}};

\draw [->] [thick] (-2.4,-1.75) arc [start angle=-90, end angle=-30, radius=40pt];
\draw (-3.4,-1.7) node [scale = .85] {\small lightcone};


\draw [thick, red] (-3.84,2) .. controls (0,-2) .. (3.84,2);
\draw [thick, red] (-3.5,2) .. controls (0, -1.3).. (3.5,2);

\draw [->] [thick]  (1,-2.3) arc [start angle=-120, end angle=-180, radius=40pt];
\draw (2.3,-2.5) node [scale = .85] {\small{$\tau =$ constant }};

\draw (0,-4.5) node [scale = 1] {\small{$g \,=\, -dt^2 + dx^2 + dy^2 + dz^2$}};


\shadedraw [dashed, thick, white](9,2) -- (13,-2) -- (17,2);
\shadedraw [dashed, thick, blue](13,-2) -- (17,2);
\shadedraw [dashed, thick, blue](13,-2) -- (9,2);

\draw [<->,thick] (13,-3.5) -- (13,2.25);
\draw [<->,thick] (8.5,-2) -- (17.5,-2);

\draw (12.65,2.5) node [scale = .85] {$t$};
\draw (17.75, -2.25) node [scale = .85] {$x^i$};
\draw (12.75,-2.25) node [scale = .85] {$\mc{O}$};


\draw [->] [thick] (14.5,2.8) arc [start angle=140, end angle=180, radius=60pt];
\draw (15.0,3.25) node [scale = .85]{\small{A Milne-like spacetime}};

\draw [->] [thick] (10.6,-1.75) arc [start angle=-90, end angle=-30, radius=40pt];
\draw (9.6,-1.7) node [scale = .85] {\small lightcone};


\draw [thick, red] (9.16,2) .. controls (13,-2) .. (16.84,2);
\draw [thick, red] (9.5,2) .. controls (13, -1.3).. (16.5,2);

\draw [->] [thick]  (14,-2.3) arc [start angle=-120, end angle=-180, radius=40pt];
\draw (15.3,-2.5) node [scale = .85] {\small{$\tau =$ constant }};

\draw (13,-4.5) node [scale = 1] {\small{$g \,=\,\Omega^2(\tau)[ -dt^2 + dx^2 + dy^2 + dz^2]$}};

\end{tikzpicture}
\end{center}
\captionsetup{format=hang}
\caption{\small{Left: the Milne universe, $a(\tau) = \tau$, sits inside the future lightcone at the origin $\mc{O}$ of Minkowsi spacetime. Right: a Milne-like spacetime, $a(\tau) = \tau + o(\tau^{1+\e})$, sits inside the future lightcone at the origin $\mc{O}$ of a spacetime conformal to Minkowski spacetime. In both cases the global coordinate system $\phi = (t,x,y,z)$ admits a past boundary for the spacetime given by the lightcone. }}\label{milne universe and milne-like figure}
\end{figure}
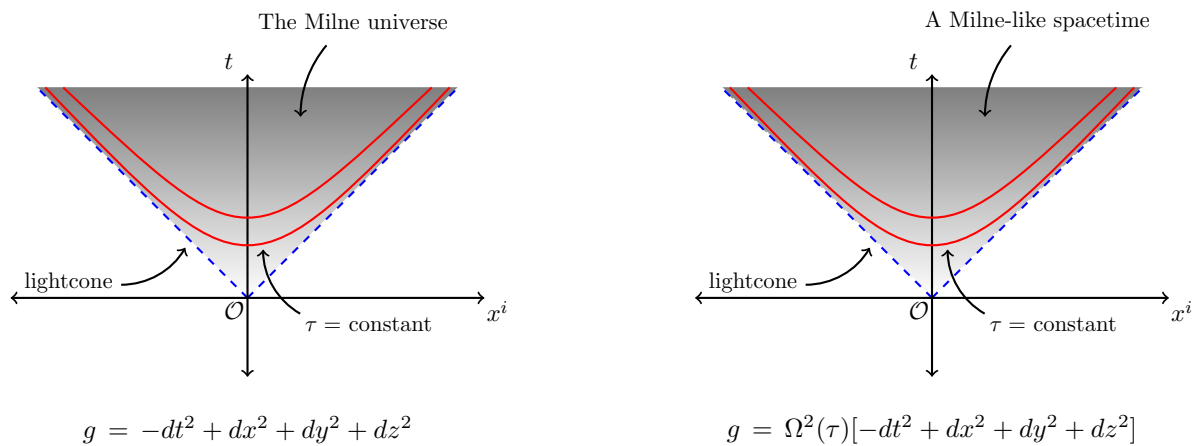

Let $(M,g)$ be the Milne universe. Then $a(\tau) = \tau$ implies $b(\tau) = \tau$ (for $\tau_0 = 1$); hence $\Omega(\tau) = 1$ identically. This shows that the global coordinate system $\phi = (t,x,y,z)$ admits a past boundary for $(M,g)$ with $(M_\ext, g_\ext)$ being Minkowski spacetime; the past boundary coincides with the lightcone $t = r$. More generally, if $(M,g)$ is Milne-like, then the proof of \cite[Thm. 3.4]{Ling_coord_sing} shows that $b(\tau) \to 0$ and $b(\tau)/a(\tau) \to 1/\tau_0$ as $\tau \to 0$ where $\tau_0$ is given in eq. (\ref{b def}). Hence $\Omega(\tau) \to \tau_0$ as $\tau \to 0$. Let $\eta$ denote the Minkowski metric. Defining $g_\ext = \tau_0^2\eta$ for points $t \leq r$ and $g_\ext = g$ for points $t > r$ gives a continuous extension $(M_\ext, g_\ext)$. (The time orientation on $M_\ext$ is determined by declaring $\pd_t$ to be future directed which agrees with the time orientation on $(M,g)$.) Therefore, as for the Milne universe, the global coordinate system $\phi = (t,x,y,z)$ admits a past boundary for $(M,g)$ which coincides with the lightcone $t = r$. See figure \ref{milne universe and milne-like figure}.

\medskip
\medskip

\section{The necessity of $a(\tau) = \tau + o(\tau)$ for past boundaries}\label{sec-necessary}

In the previous section, we saw that the global coordinate system $\phi \colon M \to \phi(M) \subset \R^4$ given by eq. (\ref{global coord}) admits a past boundary for Milne-like spacetimes. By definition, the scale factor for a Milne-like spacetime satisfies $a(\tau) = \tau + o(\tau^{1+\e})$ for some $\e > 0$. Note that this implies that $a(\tau) = \tau + o(\tau)$. In this section, we show that this condition on $a(\t)$ is necessary for an extension provided $\lim_{\tau \to 0}a'(\tau)$ exists. First, we generalize the properties necessary for $\phi = (t,x,y,z)$ to admit a past boundary for $M$.

\medskip

\begin{lem}\label{extension lem}
Let $(M,g)$ be a $C^0$ hyperbolic FLRW spacetime satisfying $a(\tau) \to 0$ as $\tau \to 0$.  Then,  $\phi(M)$ is given by eq.  \emph{(\ref{coincides eq})} and $\phi$ admits a past boundary for $(M,g)$ if and only if $b(\tau)/ a(\tau)$ converges to a nonzero finite number as $\tau \to 0$.
\end{lem}

\proof
Suppose $b(\tau)/a(\tau) \to c \in (0, \infty)$ as $\tau \to 0$. Then there exists a $\delta > 0$ such that $|b(\tau)/a(\tau) - c| < 1$ for all $0 < \tau < \delta$. Therefore $(c-1)a(\tau) < b(\tau) < (c + 1)a(\tau)$ for all $0 < \tau < \delta$.  Since $a(\tau) \to 0$ as $\tau \to 0$, the squeeze theorem implies that $b(\tau) \to 0$ as $\tau \to 0$. Consequently, $\phi(M)$ is given by (\ref{coincides eq}). Thus, by eq. (\ref{conformal metric intro eq}), a continuous extension, $(M_\ext, g_\ext)$, of $(M,g)$ is given by defining $g_\ext = (1/c^2)\eta$ for points $t \leq r$ and $g_\ext = g$ for points $t > r$, where $\eta$ is the Minkowski metric. Consequently, $\phi$ admits a past boundary for $(M,g)$ with the past boundary coinciding with the lightcone $t = r$. 

Conversely, suppose $\phi(M)$ is given by (\ref{coincides eq}) and $\phi$ admits a past boundary for $(M,g)$. By definition there is a continuous extension $(M_\ext, g_\ext)$ of $(M,g)$ with $\phi(M) \subset M_\ext \subset \R^4$. Define $C = \{(t,x,y,z) \mid t =r\}$ and $D = \{(t,x,y,z) \mid t^2 - r^2 = \lim_{\tau \to \tau_{\rm max}}b^2(\tau)\}$. (If $b^2(\tau) \to \infty$ as $\tau \to \tau_{\rm max}$, then take $D = \emptyset$.) Since $M_\ext \subset \R^4$, the topological boundary of $\phi(M)$ within $M_\ext$ satisfies $\pd \phi(M) \subset C \cup D.$
Since $\phi$ admits a past boundary for $(M,g)$, there is a point $p \in C \cup D$ and a future directed timelike curve $\g \colon [0, 1] \to M_\ext$ with $\g(0) = p$ and $\g\big((0,1]\big) \subset \phi(M)$. Therefore $p \notin D$ since continuous extensions preserve time orientations. Thus $p \in C$. Let $g_{\mu\nu}$ denote the components of the metric with respect to the coordinates $(t,x,y,z)$. Since we have a continuous extension, there is a Lorentzian metric $g_{\mu\nu}(p)$ for $T_pM_\ext$ such that $g_{\mu\nu}(q) \to g_{\mu\nu}(p)$ as $q \to p$. Applying this to points $q \in M$, it follows that $g_{xx}(p) = g_{yy}(p) = g_{zz}(p) = -g_{tt}(p) > 0$ and all other components are zero. Put $k = \sqrt{g_{xx}(p)}$ and $\Omega(\tau) = a(\tau)/b(\tau)$. We want to show $\Omega(\tau) \to k$ as $\tau \to 0$. Seeking a contradiction, suppose this is not the case. Then there is an $\e > 0$ and a sequence of points $\tau_n \to 0$ such that $|\Omega(\tau_n) - k|\geq  \e$. For this choice of $\e$, there is a neighborhood $U \subset M_\ext$ of $p$ such that $|\sqrt{g_{xx}(q)} -k| < \e$ for all $q \in U$. Since $b(\tau) \to 0$ as $\tau \to 0$ and the hyperboloids $H_\tau = \{(t, x, y, z) \mid t > 0 \text{ and } t^2 - r^2 = b^2(\tau)\}$ foliate $\phi(M)$, there is an $n$ large enough such that $H_{\tau_n}$ intersects $U$. For this choice of $n$, let $q \in H_{\tau_n} \cap U$. Then we have $\sqrt{g_{xx}(q)} = \Omega(\tau_n)$. This gives the contradiction $\e \leq |\Omega(\tau_n) - k| < \e$. 
\qed

\medskip
\medskip

The proof of (1) in Theorem \ref{thm-main} now follows from the following proposition.

\medskip
\medskip

\begin{prop}\label{main prop}
Let $(M,g)$ be a $C^1$ hyperbolic FLRW spacetime satisfying $a(\tau) \to 0$ as $\tau \to 0$. Suppose $\phi(M)$ is given by \emph{(\ref{coincides eq})} and $\phi$ admits a past boundary for $(M,g)$.  If $\lim_{\tau \to 0}a'(\tau)$ exists, then $a(\tau) = \tau + o(\tau)$. 
\end{prop}

\proof
 Lemma \ref{extension lem} implies that $b(\tau)/a(\tau)$ converges to a nonzero finite number as $\tau \to 0$. We have
\begin{equation}\label{ln b/a eq}
\ln\left(\frac{b(\tau)}{a(\tau)}\right) \,=\, \int_{\tau_0}^\tau \frac{1}{a} - \ln\big(a(\tau)\big) \,=\,   \int_{\tau_0}^\tau \frac{1-a'}{a} + \ln\big(a(\tau_0)\big).
\end{equation}
Consequently, $b/a$ converges to a nonzero finite number as $\tau \to 0$ if and only if the \linebreak improper integral $\int_{0}^{\tau_0}\frac{1-a'}{a} := \lim_{\tau \to 0}\int_{\tau}^{\tau_0}\frac{1-a'}{a}$ satisfies $\int_{0}^{\tau_0}\frac{1-a'}{a} \in \R$.

Suppose $\lim_{\tau \to 0}a'(\tau) = k \in [-\infty, \infty]$. An application of l'H{\^o}pital's rule \cite[Thm. 5.13]{Rudin} shows that
\begin{equation}\label{C^1 eq}
\lim_{\tau \to 0} \frac{a(\tau)}{\tau} \,=\, \lim_{\tau \to 0}\frac{a'(\tau)}{1} \,=\, k.
\end{equation}
Now we show $k = 1$. Since $a(\tau) > 0$, it follows that $k \in [0, \infty]$. First, consider $0 \leq k < 1$. Set $\e = (1-k)/2$. There is a $\delta > 0$ such that $a(\tau) < (k + \e)\tau$ and $a'(\tau) < k + \e$ for $0 <\tau < \delta$. Therefore 
\[
\int_0^\delta \frac{1-a'}{a} \,>\, \int_0^\delta \frac{1 - (k + \e)}{(k + \e )\tau}d\tau \,=\, \left(\frac{2}{k + 1}-1\right)\int_0^\delta \frac{1}{\tau}d\tau \,=\, \infty,
\]
which contradicts $\int_0^{\tau_0}\frac{1-a'}{a} \in \R$. Similarly, if $1 < k \leq \infty$, one obtains a similar contradiction. Thus $k = 1$. Then eq. (\ref{C^1 eq}) implies $a(\tau) = \tau + o(\tau)$.
\qed

\medskip
\medskip

We briefly allude to why the assumptions in Proposition \ref{main prop} are not optimal. Let $\bar{a} \colon [0,\infty) \to [0,\infty)$ be the continuous extension of $a(\tau)$ defined by $\bar{a}(0) = 0$ and $\bar{a}(\tau) = a(\tau)$ for $\tau > 0$. The proof of Proposition \ref{main prop} shows that $\bar{a}$ is $C^1$ and $\bar{a}'(0) = 1$. However, this is more than necessary to obtain $a(\tau) = \tau + o(\tau)$. It would have been sufficient to simply show $\ov{a}'$ is differentiable at $\tau = 0$ and $\ov{a}'(0) = 1$. 
For example, consider the scale factor $a(\tau) = \tau + \tau^2\sin(1/\tau)$. Clearly $a(\tau) = \tau + o(\tau)$ and hence $\ov{a}'(0) = 1$. Moreover, $a(\tau) = \tau + o(\tau^{3/2})$ so the corresponding hyperbolic FLRW spacetime is Milne-like, and thus $\phi$ admits a past boundary. However, $\ov{a}(\tau)$ is not $C^1$ at $\tau = 0$ since $a'(\tau)$ diverges as $\tau \to 0$ even though $\ov{a}'(0) = 1$.

 Proposition \ref{main prop} raises a natural question:
\begin{equation}\label{a' exist quest}
\emph{\text{Under what conditions does }} \lim_{\tau \to 0}a'(\tau) \emph{\text{ exist? }} 
\end{equation}
We will explore this question for the remainder of this section.

By equation (\ref{ln b/a eq}), we are motivated to define the continuous function $p\colon (0, \tau_0] \to \R$ via
\begin{equation}\label{p def}
p(\tau) \,=\, \frac{1-a'(\tau)}{a(\tau)},
\end{equation}
and we see that $b(\tau)/a(\tau)$ converges to a nonzero finite number if and only if $p$ is integrable, $\int_0^\tau p \in \R$. Note that this is an improper integral, i.e. $\int_0^\tau p := \lim_{\tau \to 0} \int_\tau^{\tau_0} p$.

Consider the following boundary value problem for $a\colon (0, \tau_0] \to (0,\infty)$:
\begin{align}\label{inv prob} 
\begin{cases}
      &a' +pa \,=\, 1, \\
      &a(\tau) \to 0 \text{ as } \tau \to 0, \\
      &p \colon (0, \tau_0] \to \R$ is continuous and $\int_0^{\tau_0} p \in \R. 
\end{cases} 
\end{align}

\noindent A $C^1$ solution to this boundary value problem is 

\begin{equation}\label{ode sol}
a(\tau) \,=\, \frac{1}{\mu(\tau)}\int_0^\tau \mu \:\:\:\: \text{ where } \:\:\:\: \mu(\tau) \,=\, e^{\int_0^\tau p}.
\end{equation}

\medskip
\medskip

\noindent\emph{Remark.} $\mu$ is well defined since $\int_0^{\tau_0} p \in \R$. To see that (\ref{ode sol}) is a solution to the boundary value problem (\ref{inv prob}), note that the fundamental theorem of calculus still holds for improper integrals; this can be verified by breaking up the integral: $\int_0^\tau p = \int_0^\e p + \int_\e^\tau p$. The same trick can be used to show that $a(\tau) \to 0$ as $\tau \to 0$.

\medskip
\medskip

 As shown in eq. (\ref{ln b/a eq}), the assumption that $b/a$ converges to a nonzero finite number is equivalent to $\int_0^{\tau_0} p \in \R$ which means that $p$ is integrable as an improper integral. If $\int_0^{\tau_0}|p| < \infty$, then we say $p$ is \emph{absolutely integrable} as an improper integral; this is the same as saying $p \in L^1$. The following lemma shows that solution (\ref{ode sol}) is unique provided we make the slightly stronger assumption that $p$ is absolutely integrable. An example of a function $p(\tau)$ that satisfies $\int_0^1 p \in \R$ but $p \notin L^1$ is $p(\tau) = \frac{1}{\tau}\sin\left(\frac{1}{\tau}\right)$.

\medskip
\medskip

\begin{lem}\label{unique lem}
Let $p \colon (0, \tau_0] \to \R$ be continuous. If $p \in L^1$, then \emph{(\ref{ode sol})} is the unique $C^1$ solution to the boundary value problem \emph{(\ref{inv prob})}.
\end{lem}

\proof
Since this is not an initial-value problem, one cannot simply invoke the standard uniqueness theorem from ODEs. Nevertheless, the proof uses similar ideas; we include it for completeness.

Suppose $a_1$ and $a_2$ are two different $C^1$ solutions to (\ref{inv prob}). Since $p \in L^1$, there is a $\tau_1 > 0$ such that $\int_0^{\tau_1} |p| < 1$. Let $f \colon [0, \tau_1] \to \R$ be the continuous function defined by $f(0) = 0$ and $f(\tau) = a_1(\tau) - a_2(\tau)$ for $\tau \in (0, \tau_1]$. 

We first show that $a_1 = a_2$ on $(0, \tau_1]$. Seeking a contradiction, suppose this is not the case. Since $|f|$ is continuous, it obtains its maximum at some $\tau_2 \in [0, \tau_1]$, and since $a_1$ and $a_2$ are different, we have $|f(\tau_2)| > 0$. 
If a $C^1$ function $a(\tau)$ satisfies $a' + pa = 1$ and $a(\tau) \to 0$ as $\tau \to 0$, then it satisfies the (improper) integral equation $a(\tau) = \int_0^\tau(1 - pa)$. Applying this to $a_1$ and $a_2$, it follows that 
\[
|f(\tau_2)| \,=\, \left|\int_0^{\tau_2}(1 - pa_1) - \int_0^{\tau_2}(1 - pa_2) \right| \,\leq\, \int_0^{\tau_2} |p(a_1 - a_2)| \,\leq\, |f(\tau_2)| \int_0^{\tau_2}  |p| < |f(\tau_2)|,
\]
which yields a contradiction. Thus $a_1 = a_2$ on $(0, \tau_1]$. 

Now we show $a_1 = a_2$ on $(0, \tau_0]$. Let $A = \{\tau \in (0, \tau_0] \mid a_1(\tau) = a_2(\tau)\}$. The above paragraph shows that $A$ is nonempty. Therefore, by connectedness, it suffices to show that $A$ is both open and closed. That $A$ is open follows from the standard existence and uniqueness theorem of ordinary differential equations. That $A$ is closed follows from continuity of $a_1 - a_2$. 
\qed

\medskip
\medskip

The following proposition provides an answer to question (\ref{a' exist quest}).

\medskip
\medskip

\begin{prop}\label{a' = 1 prop}
Let $(M,g)$ be a $C^2$ hyperbolic FLRW spacetime satisfying $a(\t)\to 0$ as $\t\to 0$.  Let $p(\tau)$ be given by eq. \emph{(\ref{p def})}. Assume one of the following two assumptions hold

\begin{itemize}
\item[\emph{(1)}] $p$ is bounded in a neighborhood of $0$, or
\item[\emph{(2)}] $\displaystyle\lim_{\tau \to 0}|p(\tau)|=\infty$ and $\displaystyle\lim_{\tau \to 0} \frac{p^2(\tau)}{p'(\tau)}$  exists.
\end{itemize}
If $p \in L^1$, then $\lim_{\tau \to 0}a'(\tau)$ exists.
\end{prop}

\proof
We show $a'(\tau) \to 1$ as $\tau \to 0$. Recall that $a(\tau)$ solves $a' + pa = 1$. Therefore, it suffices to show
\[
\lim_{\t\to 0}p(\t)a(\t) \,=\, 0.
\]
If $p(\t)$ is bounded in a neighborhood of 0, then the above limit follows from the squeeze theorem. 

Now suppose $(2)$ holds. By Lemma \ref{unique lem}, we have
\begin{equation}\label{lim ap}
\lim_{\t\to 0} p(\tau)a(\t) \,=\, \lim_{\t\to 0} p(\t)e^{-\int_0^\tau p}\int_0^\tau e^{\int_0^t p}dt.
\end{equation}
Using l'H{\^o}pital rule's, we have
\begin{equation}\label{lim l'hosp}
\lim_{\t\to 0} \frac{\int_0^\tau e^{\int_0^t p}dt}{p^{-1}(\t)}\,=\,-\lim_{\t\to 0} \frac{ e^{\int_0^\t p}}{p'(\t)p^{-2}(\t)}.
\end{equation}
From assumption, we have
$$
\lim_{\t\to 0} \frac{p^2(\t)}{p'(\t)} \,=\, c,
$$
for some $-\infty\leq c\leq \infty$. We show that integrability of $p$ implies that $c=0$. Indeed,  another application of l'H{\^o}pital's rule gives
$$
c\,=\,\lim_{\t\to 0} \frac{p(\t)}{p'(\t)p^{-1}(\tau)}\,=\,\lim_{\t\to 0} \frac{\int_0^\t p}{\ln|p(\tau)|}\,=\,0.
$$
Thus
\begin{equation}\label{lim of p^2/p'}
\lim_{\t\to 0} \frac{p^2(\t)}{p'(\t)}\,=\,0.
\end{equation}
Combining (\ref{lim ap}), (\ref{lim l'hosp}), and (\ref{lim of p^2/p'}), we have $p(\tau)a(\tau) \to 0$ as $\tau \to 0$. This completes the proof.
\qed

\medskip
\medskip

\noindent\emph{Remark.} The assumption $p \in L^1$ was only used to ensure uniqueness of solution (\ref{ode sol}) to the boundary-value problem (\ref{inv prob}). Without this assumption, we would have not been able to guarantee the equality in eq. (\ref{lim ap}).

\medskip
\medskip

\noindent\emph{Example.} Let $p(\tau) = \frac{1}{\sqrt{\tau}}$. Then $p \in L^1\big([0,1]\big)$ and satisfies (2) in Proposition \ref{a' = 1 prop}. The solution to the boundary value problem (\ref{inv prob}) is $a(\tau) = \frac{1}{2}e^{2\sqrt{x}} - \sqrt{x} -\frac{1}{2}$. Expanding $a(\tau)$ around $\tau = 0$ shows that $a(\tau) = \tau + o(\tau)$.

\medskip
\medskip

The following corollary generalizes the above example and summarizes our treatment of the boundary-value problem within the context of finding a continuous spacetime extension through $\tau = 0$.

\medskip
\medskip

\begin{cor}
Assume the hypotheses of Proposition \emph{\ref{a' = 1 prop}}. Then $\phi(M)$ is given by \emph{(\ref{coincides eq})}, $\phi$ admits a past boundary for $(M,g)$, $a(\tau)$ is given by \emph{(\ref{ode sol})}, and lastly $a(\tau) = \tau + o(\tau)$. 
\end{cor}

\proof
If $p \in L^1$, then clearly $\int_0^{\tau_0} p \in \R$. Therefore $b(\tau)/a(\tau)$ converges to a nonzero finite number as $\tau \to 0$. By Lemma \ref{extension lem}, it follows that $\phi(M)$ is given by (\ref{coincides eq}) and $\phi$ admits a past boundary for $(M,g)$. Lemma \ref{unique lem} implies that $a(\tau)$ is given by (\ref{ode sol}). Since $\lim_{\tau \to 0}a'(\tau)$ exists, Proposition \ref{main prop} implies $a(\tau) = \tau + o(\tau)$. 
\qed

\medskip
\medskip
 We make some final remarks on what happens when the hyperbolic slices have sectional curvature $-\k^2$ instead of  $-1$. In this case, the metric is $-d\tau^2 + a^2(\tau) h/\kappa^2$ where $h$ is the hyperbolic metric with sectional curvature $-1$. Our theorems in this section apply to the function $\wt{a}(\tau) = a(\tau)/\kappa$. Consequently, we find that the scale factor must satisfy $a(\tau) = \kappa\tau + o(\tau)$. Similarly, if we restore the speed of light $c$ in the metric, then we find that $a(\tau)$ satisfies $a(\tau) = c\k \tau + o(\tau)$.

\section{$a(\tau) = \tau + o(\tau)$ is not sufficient for past boundaries}\label{sec-not suff}

In the previous section, Proposition \ref{main prop} shows that $a(\tau) = \tau + o(\tau)$ is necessary for $\phi$ to admit a past boundary for $(M,g)$ provided $a(\tau)$ is $C^1$ and $\lim_{\tau \to 0}a'(\tau)$ exists. In this section we show that $a(\tau) = \tau + o(\tau)$ is not sufficient for $\phi$ to admit a past boundary. But first we show that it is sufficient for the hyperboloids to foliate $\phi(M)$ all the way down to the lightcone.

\medskip
\medskip

\begin{prop}
Let $(M,g)$ be a $C^0$ hyperbolic FLRW spacetime. If $a(\tau) = \tau + o(\tau)$, then $\phi(M)$ is given by \emph{(\ref{coincides eq})}.
\end{prop}

\proof
It suffices to show $b(\tau) \to 0$ as $\tau \to 0$. Since $b(\tau) = e^{\int_{\tau_0}^\tau \frac{1}{a}}$, it suffices to show $\int_0^{\tau_0} \frac{1}{a} = \infty$. This follows from $a(\tau) = \tau + o(\tau)$ by a simple $\e$-$\delta$ argument.
\qed

\medskip
\medskip

Now we show $a(\tau) = \tau + o(\tau)$ is not sufficient for $\phi$ to admit a past boundary. The proof of (2) in Theorem \ref{thm-main} follows from the following proposition.

\medskip
\medskip

\begin{prop}\label{not suff prop}
Let $(M,g)$ be the $C^\infty$ hyperbolic FLRW spacetime with $a(\tau) = \tau - \tau/\ln(\tau)$ for $\tau \in (0,1)$. Then $a(\tau) = \tau + o(\tau)$ but $\phi$ does not admit a past boundary for $(M,g)$. 
\end{prop}

\proof
That $a(\tau) = \tau + o(\tau)$ follows from  $-1/\ln(\tau) \to 0$ as $\tau \to 0$.  Fix $\tau_0 \in (0,1)$. Then 
\[
\ln\big(b(\tau)\big) \,=\, \int_{\tau_0}^{\tau} \frac{1}{t-\frac{t}{\ln t}}dt \,=\, \ln\big[\t(1-\ln\t)\big]-\ln\big[\t_0(1-\ln\t_0)\big].
\]
It follows that
\begin{align*}
	 \ln\(\frac{b(\tau)}{a(\tau)}\right) \,&=\,	-\int_{\tau}^{\tau_0} \frac{1}{t-\frac{t}{\ln t}}dt-\ln\(\t-\frac{\t}{\ln \t}\right)\\
	 &=\, \ln|\ln\t|-\ln\big[\t_0(1-\ln\t_0)\big].
	\end{align*}
Thus $b(\tau)/a(\tau) \to \infty$ as $\tau \to 0$. Now Lemma \ref{extension lem} implies that $\phi$ does not admit a past boundary for $(M,g)$.
\qed

\medskip
\medskip

Milne-like spacetimes satisfy $a(\tau) = \tau + o(\tau^{1+\e})$ for some $\e > 0$. In \cite[Thm. 3.4]{Ling_coord_sing} it was shown that for this scale factor, $b(\tau)/a(\tau)$ converges to a nonzero finite number as $\tau \to 0$ and so $\phi$ admits a past boundary for Milne-like spacetimes by Lemma \ref{extension lem}. The following proposition shows that the $\e$ is not necessary to achieve a past boundary and proves (3) in Theorem \ref{thm-main}.

\medskip
\medskip

\begin{prop}\label{eps not nec prop}
Let $(M,g)$ be the $C^\infty$ hyperbolic FLRW spacetime with scale factor $a(\tau) = \tau + \tau e^{-\sqrt{|\ln \tau |}}$ for $\tau \in (0,1)$. Then
\[
a(\tau) \,=\, \tau + o(\tau) \:\:\:\: \text{ but } \:\:\:\: a(\tau) \neq \tau + o(\tau^{1+\e}) \text{ for any } \e > 0,
\] 
and $\phi$ admits a past boundary for $(M,g)$.
\end{prop}

\proof
Set $f(\tau) = e^{-\sqrt{|\ln \tau|}}$.
That $a(\tau) = \tau + o(\tau)$ follows from $f(\tau) \to 0$ as $\tau \to 0$. To show $a(\tau) \neq \tau + o(\tau^{1+\e})$ for any $\e > 0$, it suffices to show $f(\tau)/\tau^\e \to \infty$ as $\tau \to 0$.  Indeed
	$$ \lim_{\tau\to 0} \left[\ln  \left(\frac{f(\tau)}{\tau^\epsilon}\right) \right] \,=\, \lim_{\tau\to 0} \left[ -\sqrt{|\ln \tau|}-\epsilon\ln\tau \right] \,=\, \lim_{\tau\to 0} \left[ \sqrt{|\ln \tau|}(-1+\epsilon\sqrt{|\ln\tau| })\right] \,=\, \infty. $$
Lastly, we want to show $b/a$ converges to a nonzero finite number as $\tau \to 0$. We have
	\begin{align*}
	 \ln\(\frac{b(\tau)}{a(\tau)}\right) \,&=\,	-\int_{\tau}^{\tau_0} \frac{1}{t+t f(t)}dt-\ln(\t+\t f(\t))\\
	 &=\, -\int_{\tau}^{\tau_0} \frac{1}{t+t f(t)}dt+\int_{\tau}^{\tau_0} \frac{1+f(t)+tf'(t)}{t+t f(t)}dt -\ln(\t_0+\t_0 f(\t_0))\\
	 &<\, \int_{\tau}^{\tau_0} \frac{f(t)+tf'(t)}{t}dt -\ln(\t_0+\t_0 f(\t_0))\\
	 &=\, \int_{\tau}^{\tau_0} \frac{e^{-\sqrt{|\ln(t)|}}}{t}dt +f(\t_0)-f(\t) -\ln(\t_0+\t_0 f(\t_0))\\
	 &=\, f(\t_0) [2\sqrt{|\ln(\t_0)|}+3]-f(\t) [2\sqrt{|\ln(\t)|}+3]-\ln(\t_0+\t_0 f(\t_0)),
	\end{align*}
	where in the second line we used the fundamental theorem of calculus, in the third the positivity of $f$ and $f'$, and in the last one integration by parts. Using l'H{\^o}pital's rule, we have
	$$\lim_{\t\to 0}f(\t)\sqrt{|\ln\t|} \,=\, \lim_{\t\to 0} \frac{\sqrt{|\ln \t|}}{e^{\sqrt{|\ln \t|}}}\,=\,\lim_{\t\to 0} \frac{1}{e^{\sqrt{|\ln \t|}}}\,=\,0.$$
	This shows
	$$\lim_{\tau\to 0} \frac{b(\t)}{a(\t)} \,<\,\infty.$$
	 Since $b/a$ is positive, it suffices to show that $b/a$ is decreasing. Indeed, computing the derivative gives $(b/a)' = -ba^{-2}(f + \tau f') < 0$.
\qed

\medskip
\medskip

\noindent\emph{Remark}. The choice of scale factors in Propositions \ref{not suff prop} and \ref{eps not nec prop} could not have been analytic. Any analytic scale factor $a(\tau) = \sum_0^\infty c_n \tau^n$ satisfying 
$a(\tau) = \tau + o(\tau)$ implies $c_0 = 0$ and $c_1 = 1$;
 hence $a(\tau) = \tau + \sum_2^\infty c_n \tau^n$. Choosing $\e = 1/2$, it's clear that $a(\tau) = \tau + o(\tau^{3/2})$ and hence the corresponding spacetime is Milne-like.

\medskip
\medskip

\section*{Acknowledgments} 
The authors are grateful to Graham Cox for posing a question at the 2022 CMS summer meeting which ultimately led to this paper. We thank Greg Galloway for helpful comments.

\medskip
\medskip
\medskip

\bibliographystyle{amsplain}

\end{document}